\documentclass[12pt,a4paper]{article}
\pdfoutput=1
\usepackage[english]{babel}
\usepackage[latin1]{inputenc}
\usepackage{amsfonts,amsbsy,bm,euscript,mathrsfs}
\usepackage{amssymb,stmaryrd,faktor,slashed}
\usepackage{color}
\usepackage[tbtags]{amsmath}
\usepackage[bookmarks=true,colorlinks=true,linkcolor=black,citecolor=black,urlcolor=black,bookmarksnumbered]{hyperref}
\usepackage[a4paper,text={170mm,257mm},centering]{geometry}
\numberwithin{equation}{section}

\def\id{\protect{{1 \kern-.28em {\rm l}}}}
\usepackage{color}
\usepackage{amsfonts}
\usepackage{amssymb}
\usepackage{epsfig}

\newcommand{\mathsym}[1]{{}}
\def\ci{\cite}

\def \bi{\bibitem}
\def \la {\label}
\newcommand{\rf}[1]{(\ref{#1})}
\def \ov {\over}
\def \ha{{\textstyle{1\ov 2}}}
\def \del {\partial}
\def \tr {{\rm tr}}
\def \G {\Gamma}

\def \ha{{\textstyle{1\over 2}}}

\def \T {{\cal T}}
\def \tr {{\rm tr}}

\def \a {\alpha}

\def \L {\Lambda}

\def \x {\xi}

\def \BI {Born-Infeld }

\def \k {\kappa}

\def \del {\partial}

\def \const {{\rm const}}

\def \s {\sigma}

\def \vp {\varphi }

\def \td {\tilde }
\def \d {\delta}
\def \ci {\cite}

\def \inv {^{-1}}
\def \ov {\over }

\def \f {{\rm F}}

\def \iffa  {\iffalse}

\newcommand{\be}{\begin{eqnarray}}
\newcommand{\ee}{\end{eqnarray}}
\def \G  {\Gamma}

\def \foot{\footnote} \def \ed {\end{document}}
\def \L  {{\bar L}}
\def \T {{\cal T}}   \def \G  {{\Gamma}}
\def \M   {{\cal   M}} \def \Z  {{\cal Z}}
\def \MM {\mu} 
\def \M  {{\cal M}}  
\def \z  {\zeta} 
\def  \k { \varkappa}
\def \MM {\k}
 \def \cG  {{\cal G}}
  \def \Li {{\rm Li}}
 \def \no {\nonumber}

\renewcommand{\title}[1]{\vbox{\center\LARGE{#1}}\vspace{5mm}}
\renewcommand{\author}[1]{\vbox{\center#1}\vspace{5mm}}

\begin{document}
\iffa 
\overfullrule=0pt
\parskip=2pt
\parindent=12pt
\headheight=0in \headsep=0in \topmargin=0in \oddsidemargin=0in
\fi

\vspace{ -3cm} \thispagestyle{empty} \vspace{-1cm}
\begin{flushright} 
\end{flushright}
 \vspace{-1cm}
\begin{flushright}  Imperial-TP-AT-2020-03
\end{flushright}




\vspace{2.5cm}

\begin{center}

{\Large\bf  Comments   on   open  string\\ \vspace{0.2cm}
 with  ``massive"  boundary term}

\vspace{1.5cm}

{
Arkady A. Tseytlin\footnote{\ Also at the Institute for  Theoretical and Mathematical Physics of  Moscow State University 
 and Lebedev Institute, Moscow. \ \ 
tseytlin@imperial.ac.uk}
}

\vspace{0.5cm}
\vspace{0.15cm}
Blackett Laboratory, Imperial College, London SW7 2AZ, U.K.

\end{center}

\vspace{0.5cm}

\begin{abstract}
We  discuss  possible  definition of  open string path integral   in the presence  of additional  boundary  couplings
corresponding  to   the presence of masses  at the ends of the string.    These  couplings are not  conformally invariant 
implying  that  as in a non-critical string case  one is to  integrate over the  1d metric or reparametrizations of the  boundary. 
We  compute the  partition function on the disc   in the presence of  an additional  constant gauge field  background 
and  comment on the structure of  the  corresponding scattering  amplitudes. 
\end{abstract}

\newpage



\setcounter{footnote}{0}
\setcounter{section}{0}

\section{Introduction}

Relativistic   string   is  a remarkably  rigid  theory:
 it  is hard  to modify it while preserving  its  quantum-mechanical  consistency  and solvability. 
 One of the key  conditions is 2d conformal invariance, the   preservation of which 
 imposes constraints  on the target space  dimension and  geometry. 
 
 There were several attempts to generalize  the open string theory by  adding masses  at the ends 
 (see, e.g.,  \cite{Chodos:1973gt,Bars:1975dd,Barbashov:1977nq, Sever:2017ylk, Sonnenschein:2018aqf}   for some old   and recent papers).
 At the classical level   adding masses  to  the  ends of  an open string corresponds   to considering the action 
\begin{align}
\la{0}   &  I= I_0 + I_\del\ , \  \qquad 
  I_0 =  T  \int d^2 \s\  \sqrt { \det h_{ab}  }  \ , \qquad h_{ab} =   \del_a x^n \del_b x_n  \ , \\
&   \la{00}   I_\del  =  m_1 \int d \tau \sqrt { - \dot x^2} \Big|_{\s=0} +  m_2 \int d \tau \sqrt { - \dot x^2} \Big|_{\s=\pi} \ . 
\end{align}
While  it   may be   used   for  developing  an effective perturbation theory   near a   long
 semiclassical string, 
this   action is  not a  good starting point for   quantization  of  short  fundamental strings  being  non-linear.

Instead, one  may    introduce   an auxiliary metric $g_{ab}$    and  a boundary metric $e$   and consider  the analog of the  Polyakov  string path integral  \cite{Polyakov:1981rd,Polyakov:1987ez}  with the 
following action defined, e.g.,    on a  disc  
\begin{align}  \la{1} &
I= I_0 + I_\del , \  \quad I_0 =\ha T  \int d^2 \s\,  \sqrt g\,   \del^a x^n \del_a x_n   \ , \qquad \\
&  I_\del =
 \ha \int^{2\pi}_0  d \vp\  \big(e\,   \T_0    +   e^{-1}  Q_0\,   \dot x^m \dot x_m       \big)   \ . \la{121} \end{align} 
Here we  use the  Euclidean  notation,  $T= { 1 \ov 2\pi \a'}$ is string tension,  $e=e(\vp)$   is an   einbein and 
$\T_0,Q_0$ are constant    parameters (one of them may  be absorbed into $e$). 
Solving for $e$  classically  gives  the  ``mass term" like in \rf{00}\foot{Strictly speaking,  disc  or 
 half-plane   with the above mass  term at the boundary  corresponds   to  an  unphysical choice of  $m_1 = - m_2=M$  when
transforming from  the action \rf{0} on  the  strip  with two boundaries and two masses at ends.  Moreover, 
in the absence of conformal invariance    descriptions   using different domains may not be equivalent. 
} 
\be \la{2} 
\hat I_\del =  M \int d \vp\ |\dot x| \ , \qquad\qquad   M\equiv  \sqrt{ \T_0  Q_0} \ .   \ee 
$I_\del$  in \rf{1}     may  be  viewed  as a special case  of the   boundary action \cite{Fradkin:1985qd}
\be \la{3} 
  I_\del =
 \ha \int^{2\pi}_0  d \vp\  \Big[   e   \T(x)  +   e^{-1}    Q _{mn}  (x)\,   \dot x^m \dot x^n  - i    A_m(x)\,  \dot  x^m   + ...       \Big] \ ,  \ee
where $\T$    may be interpreted   as a  condensate of an   open string tachyon, $Q_{mn}$  -- 
of a  spin 2  massive    open string mode,  $A_m$  -- of  a massless  vector field,   etc. 

 In general,  for   fixed  $e$  the  presence of   non-trivial 
$\T, Q, ...$     couplings breaks scale  invariance (beta-functions  for $\T,Q, ...$ will be non-zero, see, e.g.,   \ci{beta}). 
As a result, if, for example,  $\T=\T_0, \ Q_{mn} = Q_0  \delta_{mn}$ as in  \rf{1}, then  
 $e$ will not  decouple, i.e.   there will be a  scale (and 1d  $SL(2)$ conformal)  anomaly. 
 This will happen even  if the 2d bulk   Weyl  anomaly cancels out  (which, of course, 
  requires  $D=26$ \cite{Polyakov:1981rd}).\foot{An example of a quantum-consistent 1d CFT  is a  string in a constant Maxwell 
 field  background  with the coupling term   $\int d \tau  F_{mn} x^m \dot x^n $. 
 One may also consider other options like non-local 1d actions that are scale invariant (cf.,  e.g.,  \ci{gr}).}

A  possible  way out is to   consider a   kind  of   ``light" version of  non-critical string theory 
  where  the  bulk conformal  factor is  decoupled (in $D=26$) 
but one is still to  integrate  over  the 1d metric  $e$. 
This   will  allow one  to 
  absorb the  1d  scale anomaly or  the corresponding  2d   UV 
divergences into a redefinition of $e$.  If  the reparametrization invariance  is     assumed to be preserved by a  regularization, 
 we may then fix a 1d reparametrization gauge   as 
$e(\vp) = L=$const  and   integrate over  the remaining constant  parameter $L \in (0,\infty) $.\foot{This is,   of course,   what is done in the familiar 
 proper-time   representation of a 1-loop effective action (say, with kinetic operator $\Delta =  - \del^2 + \T(x) + ..$),   i.e. 
    $\Gamma_1=\ha \log \det \Delta 
 = - {1 \over 2} \int {d L\ov L}  \tr(  e^{- L \Delta} )$.}

 Then the string partition function  on  a   disc  will    be given by 
 \begin{align} \la{4}
& \Z= \int^\infty_0   d L\  \mu(L)  \ e^{ - \pi    L  \T_0}  \ \hat \Z \ , \\ \la{44}
 &\hat \Z=  \int [d x]      \exp \Big[ - I_0  -  \ha  \int^{2 \pi}_0  d \vp\  \big(      L^{-1}  Q_0    \dot x^m \dot x^m    - i A_m(x)  \dot x^m + ...  \big)  \Big]   \ . 
 \end{align} 
  In  the standard   critical     string theory  (where $\T_0=Q_0=0$)  
     the integral over $L$    should   decouple  (and,   in fact, it should be  removed   as a result of dividing over the 
      Weyl   gauge group on the disc). 
        One  may expect that  the measure 
      $\mu(L) $   should be proportional to   $    L^{ k (D-26)}$  and  should  thus     
     be trivial  in the critical dimension.  
 The precise     form of $\mu(L)$ remains one of the 
        open  questions.\foot{The definition of the measure   is   subtle  as  it  requires,  in
particular, a   proper choice of the boundary condition for the  bulk conformal
factor  relating it to $e$. The present  case is different from the Wilson loop case \ci{alv}
as  the boundary conditions are different -- for $M=0$ we should recover the Neumann  boundary condition, not the Dirichlet one.} 
In general, the  gauge fixing procedure leading to \rf{4} 
in the  case of a  disc topology is  non-trivial and deserves further  investigation.\foot{We thank the referee for 
pointing out  some related  issues.}

        Another issue   will be how to define   the corresponding vertex  operators and thus scattering amplitudes. 
In general,  the  corresponding  Green's function will  depend on 
the metric  $e$  and thus  there  will be additional $L$-dependent terms reflecting 
 breaking of  1d scale invariance.  For  constant    gauge field   background 
 $F_{mn}=\const$   these are only   linear  divergences  (that can be absorbed into  tachyon coupling) 
   but,   in general, 
 there will be also  non-trivial log divergences.
   If   the condition of decoupling of $e=L$ is not  satisfied  we cannot use the usual 
     marginality condition to determine the vertex operators.  
     A naive  guess   is that the vector field vertex operator    which does not  directly 
   couple  to $e$    may  still remain the same.\foot{The condition of marginality  of  the vector vertex operator will 
    be sensitive  to $e$  via modified   Green's function so  it is unclear why  it should remain to represent a massless particle.}
   
    Below  in section  2 we shall   first consider the   disc partition function  in  the  abelian   constant  gauge field strength    background 
    and then in section 3    make comments on the scattering amplitudes.

  \section{Partition function in constant  gauge field   background}
  
To get an idea of how the  integrand of the  $L$-integral   in   \rf{4}   may look like  
  let us consider  the  generating functional for scattering of   soft photons  
which in the standard     critical    string theory   is described  by  the Born-Infeld  action.
 It is   given   by  the  disc  partition function  in constant  strength  background for the  vector 
field  in \rf{3}, i.e. $A_m = -\ha  F_{mn}  x^n$.  We shall  use   the  discussion in  Appendix  of  \cite{Tseytlin:1999dj}.

 Integrating over the values of  the string coordinates at the internal points
 of the disc  we  get  the following expression for $\hat \Z$ in  \rf{44}  
 \begin{align}  \la{5}  \hat \Z=c_0 \int d^{D} x_0 \  Z \ , \qquad \qquad 
 Z =  \int [d\xi] \ e^{-\hat  I_{\del}} \ , \qquad  \ \ \ c_0 \sim T^{-D/2} \,   \\ \la{6}
\hat I_{\del} = 
\ha  \int^{2\pi}_0  d \vp\  \Big(  T  \x^m G^{-1}  \x^m   + 
 \M  \dot \x^m \dot \x^m   +   i F_{mn} \x^n \dot \x^m   \Big)  \ ,  \ \qquad \M\equiv  L^{-1} \, Q_0 \ .  \end{align}
Here  $\hat I_{\del}$   is      the  effective action 
  at the boundary of the disc  and   
we  isolated the constant zero mode  in $x^m = x_0^m+\xi^m(\vp) $, \ \ 
 $\x^m(\vp)  ={1\ov \sqrt \pi}  \sum^\infty_{n=1} (a^m {\rm cos}\ n \vp 
 + b^m {\rm sin}\ n \vp)$. The  
 scale-invariant   non-local operator $G^{-1}$ in \rf{6}   is the 
inverse of the  restriction of the  Neumann  function on the 
disc to its boundary\foot{Note that the $\delta$-function  on non-constant $\xi$ is   $\bar \delta (\vp) = {1\ov \pi}\sum^\infty_{n=1} \cos n \vp $.}
\be \la{7}
G(\vp_1,\vp_2) = {1\ov \pi}\sum^\infty_{n=1} 
{1  \ov n} \cos n \vp_{12}   \ , \ \ \ \ \ \ \ \ 
G\inv (\vp_1,\vp_2) = {1\ov \pi}\sum^\infty_{n=1} 
 { n} \cos n \vp_{12}   \ , \ \ \ \  \vp_{12} =\vp_1 - \vp_2  \ee 
The action  thus contains  the effectively  ``first-order" term 
($\sim T$)  and  the second-order  term ($\sim \M$)  in 
1d derivatives  and  interpolates between the  standard  massless
string theory case $T\not=0, \ \M=0 $
  and the  standard particle  case   $T=0, \ \M\not=0$
 which  appears in the  Schwinger computation of  $\log \det [-D^2(A)]$ in constant $F_{mn}$ background \cite{Schwinger:1951nm}. 
The resulting partition function will then  interpolate between the \BI  (string)
and  the  Schwinger (particle) 
expressions.

Putting  $F_{mn}$ into  the  block-diagonal form      and   concentrating  first on a single 
 $(1,2)$ block  
we find  after  integrating over the  coordinates $\x^1,\x^2$ 
\begin{align}  \la{8}  &Z_{12}  =    Z_{12} (\M)\,  Z_{12} (F,\M)  \ , \\
\la{9}   &Z_{12} (\M) \sim \prod_{n=1}^\infty (Tn + \M n^2)^{-2}
\sim\   \M \ \Big[\prod_{n=1}^\infty (1 + { T\M\inv \ov n} )\Big]^{-2} \ , \\
\la{10}  &Z_{12} (F, \M) = 
\prod_{n=1}^\infty\Big[  1 + {\f^2 \ov (T + \M n)^2 }  \Big]^{-1} \ ,  \ \ \ \ \qquad  \f \equiv F_{12} \ . \end{align} 
$Z_{12} (F,\M)$   in \rf{10}  depends only on the  dimensionless  ratios
$T\inv \f$ and $T\inv \M$. 
We shall ignore  the (power divergent) $F$-independent  factor   $Z_{12} (\M)$
which can be absorbed into the renormalization
of tachyon  coupling $\T_0$  in \rf{4}.\foot{Recall  that  according to \rf{6}   we have 
$\M=L^{-1} Q_0 $  so that  $\M$ depends on $L$.
All  linear divergences should be renormalized away;  this is also part of  subtracting the $SL(2,R)$  Mobius 
 volume  in the standard open string  theory set-up \ci{mob}.}

When $\M=0$  the  factors in the product in $Z_{12} (\f,\M)$ are
 $n$-independent, and  using the   standard  regularization prescription 
$\prod_{n=1}^\infty c =  c^{-1/2}$  
 (with the  linear divergence again absorbed into the  tachyon coupling)
  we get the  familiar   Born-Infeld  expression 
  \be  \la{11}
  Z_{12}(F,\M=0)  =\sqrt{ 1 + (T\inv\f)^2}  \ . \ee
  For  $T=0$ we get  instead  the  Schwinger expression 
$Z_{12}(F,\M)\big|_{T=0}  = { \pi \M\inv \f \ov \sinh (\pi \M\inv \f)}$.

In general, for $\M\not=0,$ \   $Z_{12} (F,\M)$  in \rf{10} 
 may be  expressed in terms of $\Gamma$-functions  as 
\be \la{12}
Z_{12} (F,\M) = { \G ( { T + \M +  i\f \ov \M}) 
  \   \G ( { T + \M -  i\f \ov \M})  \over
[ \G ({ T + \M   \ov \M}) ]^2 }=\Big| { \G ( { T + \M +  i\f \ov \M})
   \over
\G ({ T + \M   \ov \M})  }\Big|^2 
 \ .  \ee
The general expression for the partition function is 
 the product of factors for each eigenvalue $\f_p$  of the field 
strength $F_{mn}$ 
\be \la{13}
Z (F,\M) = \prod^{D/2}_{p=1} \bigg[
  { \G ( { T + \M + i\f_p \ov \M} )\ 
  \G ( { T + \M -  i\f_p \ov \M})  \over
[ \G ({ T + \M   \ov \M}) ]^2 } \bigg] 
 \ . \ee
Once again,   in the ``point-particle"  limit  $T\to 0$  we get the standard Schwinger expression \cite{Schwinger:1951nm}
\be Z(F,\M)\Big|_{T\to 0} \to \prod^{D/2}_{p=1} { \pi  \M\inv
 \f_p \ov \sinh ( \pi  \M\inv \f_p )} \ . \la{210}  \ee
Taking  the limit $\M\to 0$  and 
using the  Stirling formula 
$\G({z \to \infty})  = \sqrt{ 2\pi \ov z}  ({z\ov e})^z [ 1 + O({1\ov z})]$
we  find  from \rf{13} 
\be  \la{14} 
Z\Big|_{\M \to 0}   = \prod^{D/2}_{p=1} 
  \sqrt{ 1 + (T\inv \f_p)^2} \ \Big[1 + (T\inv \f_p)^2\Big]^{\M^{-1} T} 
\Big({ 1 + iT\inv \f_p \ov   1 -  iT\inv \f_p   }\Big)^{i\M^{-1} \f_p}\  \big[ 1 +  O(\M) \big] \ . 
 \ee
The   mass parameter $\M = L^{-1} Q_0 $ thus plays here the  role  of a   UV  cutoff  with linear  divergences  proportional to $L$. 
Eq.  \rf{14}  then  reduces to the  standard  \BI expression  times  an extra  divergent factor
\be\la{15}
  Z\Big|_{\M \to 0}  =  \sqrt{ \det(\d_{mn} +  T\inv F_{mn})} \ \ e^{ { \M^{-1}  } f(F)} \ ,  \ee
  that can be  ``renormalized"  away   by absorbing it  into the $\T_0$ term in \rf{4}
  that also scales  linearly with  $L$.

Finally, it remains to substitute the expression for $ \hat \Z$ in \rf{5},\rf{13}   into \rf{4}  and integrate over $L$. 
Let us assume for simplicity  that we have just one    magnetic field component $F_{12}=\f$    and  plug \rf{12} into 
the integral over $L$  in \rf{4} 
\begin{align} 
\la{20} &\Z  =  c_0  V_D  ( 1 +\bar \f^2)   \int^\infty_0   d L\  \mu(L)  \ e^{ - \pi  T^{-1} M^2   \L  }\       H(\L, \bar  \f) \ ,  \\
  \la{21}
& H(\L, \bar \f ) \equiv    { \G \big(  \L  ( 1  + i \bar  \f) \big) \ \G \big(  \L  ( 1  - i \bar \f) \big) 
    \over
[ \G ({ \L }) ]^2 } \ ,\qquad \qquad 
  \L \equiv  T Q_0^{-1}  L \ , \ \ 
  \ \ \     \bar \f \equiv   T^{-1} \f \ . 
   \end{align}
   Here $V_D$ is the  volume factor in \rf{5}  and 
we used  the definition $M^2=\T_0 Q_0 $ in \rf{2}. 
The standard  massless string theory limit corresponds to  $Q_0\to 0$, $M^2\to 0$ or $\bar L \to \infty$ for fixed $T$. 
This is also  the limit $\M  \to 0$ in \rf{14}.   
If the measure $\mu(L) \sim L^\gamma$ with $ \ \gamma > 0$  then  the  resulting integral over $L$  is regular
and gives a finite expression for the partition function 
as a function of the tension $T$, magnetic field $\f$ and  the mass parameter $M$.

\section{Scattering amplitudes}

Next, we may look at the generalization of the standard   vector scattering amplitudes 
to the case of  non-zero  mass parameter $M$ in \rf{121},\rf{2}.  
 To compute  the  scattering amplitudes   we need to specify   (i)  vertex operators, 
 (ii)   the modified ($L$-dependent)   boundary Green's function, and 
 (iii)   integrate over $L$ as in \rf{4}.

Let us first   comment on  the Green's   function. 
To find \rf{7} we followed  \cite{Fradkin:1985qd}   and started with the Neumann     function on the disc. 
This does not restrict the boundary value  of the string coordinate and   just amounts to integrating out 
 its values  in internal points of the disc. 
 For example,  if one has boundary coupling to an external vector, it   classically  modifies  the boundary conditions 
 and  that leads to an  $F$-dependent  Green's function    \cite{Abouelsaood:1986gd}. However,  the same result is obtained 
  by restricting the Neumann function  to the boundary and then considering  the purely boundary theory  as  in \rf{6}. 
 Explicitly,  if  $z=  r e^{i \vp}$  ($ 0 < r < 1$)  is a coordinate on a disc  then 
\be \la{22}
N(z,z') = - { 1 \ov 2 \pi} \big(   \log |z-z'|  + a \log |z - \bar z'^{-1}| \big) \ , \qquad \qquad 
G(\vp, \vp')  \equiv  N( e^{i\vp}, e^{i\vp'}) \ , 
\ee
where $a=1$   corresponds to  the Neumann function and $a=-1$   to  the Dirichlet  function. 
Then the  boundary value $G(\vp, \vp') $  is  (for $a=1$) 
\begin{align} \la{23}
&G(\vp, \vp') \equiv G(\vp-\vp') =  -   { 1 \ov 2 \pi}  \log \big[2-2 \cos( \vp-\vp')\big] = {1\ov 2 \pi}  \sum^\infty_{n=1} 
{1  \ov n} \cos n ( \vp-\vp')
\ , \\
\la{203}
&G^{-1}(\vp) =  - {d^2 \ov d \vp^2}   G(\vp) = -   { 1 \ov 4 \pi}   { 1 \ov \sin^2{ \vp-\vp' \ov 2}} =
{1\ov  \pi}  \sum^\infty_{n=1} 
{ n} \cos n ( \vp-\vp')\ , 
\end{align} 
where we used that $\log (1 + b^2 - 2 b \cos \a) =- 2 \sum^\infty_{n=1}  { b^n\ov n} \cos n \a$.

  Integrating over the string fluctuations inside the disk 
  we get 
 the boundary action \cite{Fradkin:1985qd}    in \rf{6}, i.e.\foot{Here   we  may not distinguish between 
 $x(\vp)$   and its non-constant  part $\x(\vp)$ as  the constant  $x_0$ drops out under the integral.} 
 \be \la{29}
\hat I_\del =  \ha  T  \int^{2\pi}_0  d \vp_1 d\vp_2 \,    \x^m (\vp_1)  G^{-1} (\vp_{12})    \x^m (\vp_2)  +
\ha \M    \int^{2\pi}_0  d \vp\,  \dot \x^m \dot \x^m   
\ .  \ee
 Since  from \rf{7} $G^{-1}(\vp) =  - {d^2 \ov d \vp^2}   G(\vp)$   we get 
  \begin{align}  \la{30}
&\hat I_\del =    \ha  T  \int^{2\pi}_0  d \vp_1 d\vp_2 \    \dot \x^m (\vp_1) 
\hat  G (\vp_{12})     \dot  \x^m (\vp_2)   = \ha T \int^{2\pi}_0  d \vp_1 d\vp_2 \     \x^m (\vp_1) 
\, \cG (\vp_{12}) \,   \x^m (\vp_2) 
\ ,   \\ 
 &\hat G(\vp) \equiv G(\vp)   + \MM\,  \bar \delta(\vp) =  {1\ov \pi}  \sum^\infty_{n=1} ( { 1\ov n }  + \MM )     \cos n \vp \ , \qquad 
  \cG(\vp) =   - {d^2 \ov d \vp^2}   \hat G(\vp)  
   \ ,  \la{31} \\
   &\la{307}
   \qquad  \MM \equiv  T^{-1}  \M=   L T^{-1}  Q_0 \ .
 \end{align} 
 Here    $\cG$  
is  the effective  boundary    Green's  function  corresponding to the action \rf{6}  containing  the ``mass term"  $\M \dot \xi^2$ 
 \be \la{24} 
\cG(\vp, \vp') \equiv \cG(\vp-\vp')=  {1\ov \pi}  \sum^\infty_{n=1}     g_n   \cos n ( \vp-\vp') \ , \quad  \ \ \ \ \ \   g_n =  {1  \ov n +  \MM\,   n^2 } 
   \ .   \ee
    $n$ in $g^{-1} _n$  comes   from $G^{-1}$ in \rf{7} and $n^2$ from   second-derivative term in \rf{6}.
 It is the determinant    of $\cG$ that appeared in \rf{9}. 
 The same   expression for the Green's function   should   be found  if one first modifies   the classical boundary conditions
   due to the presence   of the boundary  mass term $\M \dot \xi^2$.

 
  While  a systematic   way  of constructing   vertex operators in the ``non-critical"  massive case remains to be found 
   we may be guided by the  fact that  at least the  vector field   couples  to the  open string  ends in a  (classically) scale invariant way. 
  Let us thus  assume 
  that we may start with    the  standard    vertex operator corresponding to the  vector coupling in \rf{4}, i.e. 
 \be\la{25}
   V(\z, p) = \int d \vp \, \z_m  (p)\ \dot \xi^m(\vp) \,    e^{i p_m \xi^m (\vp) } \ .    \ee
 The generating functional   for  correlators  of unintegrated   vertex operators    (computed now with the 
 Green's function in   \rf{24})     may   be written as (cf. \rf{4},\rf{5},\rf{6}) 
 \begin{align}
  \la{26}
 &Z(\z,p) =  \int^\infty_0   d L\  \mu(L)  \ e^{ - \pi    L  \T_0}    W \ , \\
 &  \la{266}
 W= 
   \int [d \xi ]    \  \exp \Big[ {   \int^{2\pi}_0  d \vp\  \big( -   \ha  T  \x^m G^{-1}  \x^m    +    \hat \z_m  \dot \xi^m   + i \hat p_m   \xi^m   \big)   } \Big] \ . 
 \end{align}
 Here $\hat  \z_m =  \sum^N_{k=1}  \z^{(k)}_m(p_k)  \delta (\vp - \vp_k) , \ \ \ 
\hat  p_m =  \sum^N_{k=1}  p^{(k)} _m  \delta (\vp - \vp_k)  $
and to find the expression for the 
 $N$-point scattering amplitude one  needs to take  the  relevant  multi-linear term in $\z^{(k)}$. 
The integral over  the  constant zero   mode   $x^m_0$  in \rf{4}  gives, as usual,  the  total  momentum conservation  delta-function. 

Doing the Gaussian integral over  $\xi^m$   gives 
\be \la{27}
W=  \exp \Big[ \ha  T^{-1} \sum^N_{k,k'=1}  \Big(  - \ddot \cG_{kk'} \z^{(k)}\cdot \z^{(k')}   + 2 i \dot \cG_{kk'} \z^{(k)}\cdot p^{(k')}    -  
    \cG_{kk'} p^{(k)}\cdot p^{(k')}  \Big) \Big]
 \ee
 where  according to  \rf{24}   we get   $  \cG_{kk'} = {1\ov \pi}  \sum^\infty_{n=1}     g_n   \cos n ( \vp_k-\vp_{k'}), \ \ \   g_n =  {1  \ov n +  \MM\,   n^2 },  $  and thus 
 \be \la{28}
\dot  \cG_{kk'} = - {1\ov \pi}  \sum^\infty_{n=1}    n  g_n   \sin n ( \vp_k-\vp_{k'}), \ \ \  \ \ \ \ \ 
\ddot   \cG_{kk'} = - {1\ov \pi}  \sum^\infty_{n=1}     n^2  g_n   \cos n ( \vp_k-\vp_{k'}) \ . 
 \ee
  This leads to the standard integrands   for the vector amplitudes on the disc  when  $\M=0$, i.e.   when  $g_n= {1\ov n}$.
 Under the  integral   over $L$ one should still have $SL(2,R)$    Mobius symmetry    so  one  may  use it to  fix  3  points. 
 One may also  choose   not  to  fix the Mobius symmetry   explicitly   -- as the  disc  Mobius  group  volume is only power divergent  \cite{mob}
    this   divergence  should also  be 
  possible to absorb into  a  renormalization of the  tachyon coupling  $\T_0$. 
  
\

Let us note that an  alternative representation  for the path integral  appeared in the context 
 of   computing 
the  Wilson loop (WL)  expectation value  (see  \ci{alv})
  with fixed target-space contour: there   the integral over 1d metric $e$  is equivalent to integrating over reparametrizations $s(t)$ 
of the  boundary ($ds= e(t) dt$). 
It is useful   to  map
the unit disk  $(r, \vp)$ onto the upper half-plane 
using 
$z= i \frac{1+r\, e^{i \vp}}{1-r\, e^{i \vp}} $;  then   the boundary at  $r=1$  is mapped onto the real axis 
$-\infty <t < +\infty$ by  $
t(\vp)= -\cot \frac {\vp}{2} $. 
The boundary restriction of the  Green's function in  \rf{7},\rf{23} 
mapped  to the  half-plane  is 
 \be 
 \la{32}   G(t) = -{1\ov \pi} \log |t| \ , \ \ \qquad \qquad  G^{-1} (t) = -{1\ov \pi\, t^2} \ . 
 \ee 
 Ignoring  the  boundary mass term   and  introducing the integral over reparametrizations one gets 
 (see  \ci{pol,Orland:2001rq,Makeenko:2009rf})
 \begin{align}
&\qquad \qquad  A[\vec x]=
\int  [d s(t)]  \ e^{- \hat I} \ ,\la{333}  \\    &\hat I= -\tfrac{1}{ 2\pi }  T
\int 
d t_1 
d t_2\,
\dot{\vec x}(t_1)\,  \log |s(t_1)-s(t_2)| \,  \dot {\vec x}(t_2)
= \tfrac{1}{ 2\pi }   T  \int d s_1 \, d s_2
{\big[\vec x(t(s_1)) - \vec x(t(s_2)) \big]^2 \ov (s_1-s_2)^2}
\label{33}
\end{align}
where the integrals go over the real line (cf. \rf{30})
and ${\vec x}$ is the  counterpart  of $\xi^m$ in \rf{29}. 
The analog of the  mass  term  is 
$I_\del = \ha \M \int dt \ \dot{\vec x}(s(t))\,  \dot{\vec x}(s(t)) $.
If we are interested,  for example, in tachyon scattering amplitudes  we  may  just insert 
   momentum-dependent  factors  and integrate over all boundary functions   $\vec x(t)$.  
   Alternatively, one   may do  Fourier transform  of the WL 
  expectation value  and  then pick up  a  step-function - like  contour for $\vec  p(t)$   \cite{Makeenko:2009rf}. 
  This    may be viewed as a particular off-shell prescription for tachyon scattering amplitudes.

 To appreciate the technical difficulty between the  $\M=0$ and $\M\not=0$ cases it is useful 
 to  find the explicit form of the   counterpart of the Green's function \rf{24}  in the  half-plane parametrization. 
 We may write $ \cG$ in  \rf{24}  as 
 \be \la{41} 
 \cG(\vp) =  {1\ov \pi}  \sum^\infty_{n=1}   { \cos n  \vp  \ov n + \MM\,  n^2}    =  G(\vp;0) - G(\vp;\MM^{-1} ) \ , \ \  \ \ \ \ \ \ 
    G(\vp; b ) \equiv   {1\ov \pi}  \sum^\infty_{n=1}   {   \cos n  \vp  \ov n +  b } \ .
       \ee
  Let us set  $w= e^{i \vp},  \ w'=e^{i \vp'}$  so that $G(\vp-\vp'; b)$   may be written as 
 \begin{align} 
    G(w,w';b ) & \equiv   {1\ov 2\pi}  \sum^\infty_{n=1}   {1 \ov n +  b }    \big({w\ov w'}\big) ^n + c.c. ={1\ov 2\pi}
    \Big[  \Phi\big( {w\ov w'}, 1, b\big)  -  b^{-1} \Big]  + c.c.\no  \\
    &={1\ov 2\pi}   \sum_{r=0}^\infty  (-b)^r  \,  \Li_{r+1}  \big( {w\ov w'}\big)   + c.c.  \ .
     \la{41y}   \end{align}
      Here 
            $\Phi(u, r, b) \equiv \sum_{k=0}^\infty   { u^k \ov ( k + b)^r}    $ 
 is the Lerch transcendent generalizing the Hurwitz  $\zeta$-function, i.e. 
 \be \la{41z} 
  \Phi(u, 1, b) -  b^{-1}  =
  \sum_{n=1}^\infty   { u^n \ov  n + b} =    \sum_{r=0}^\infty  (-b)^r   \Li_{r+1} (u) \ , 
  \ee 
  where $\Li_s(u)  
  = \sum^\infty _{k=1} {u^k\ov k^s} $ is the polylogarithm function.  
  Note that 
  $ 
  \sum_{n=1}^\infty   { u^n \ov  n + b}  =  z \Phi(u, 1, 1+b ) .$
  
  The map  from the disc  to  half-plane  is 
      $z= i \frac{1+r\, e^{i \vp}}{1-r\, e^{i \vp}}$   
       or   
       \be \la{50}
       w= {  z -i \ov   z + i } \ee
         with    $|w|=1$ if $z$ is real. 
       The   massless  boundary  Green's function  expressed   in terms of  real $z,z'$  or $|w|=|w'|=1$   is 
    \be \la{53}
 G(w,w';0)  =   -  \tfrac{1}{ \pi}\log | 1 -  \tfrac{w}{ w'}|  = -   \tfrac{1}{ \pi}   \log | w -  {w'}|=
    -   \tfrac{1}{ \pi}\log \tfrac{ 2 |  z-z' | }{ |z+i|  |z' +i| } \to   -   \tfrac{1}{ \pi}\log  |   z-z' |, 
    \ee    
         where we used  the freedom in the Green's function     
        $ G \to G + f(z)  + \td f  (z')$  to  get the standard  expression (cf. \rf{32}) 
        at the boundary of half-plane. 
         Then  $  \cG(z,z') $ can be    found directly from \rf{41y},\rf{50}. 
              For example, the small $\MM$ expansion  of $\cG$ in \rf{41} is  
       \be \la{54} 
       \cG=   -  {1\ov   \pi}\log | 1 -  {w\ov w'}|   +  {1\ov  2 \pi}\Big[ \MM  ( {w\ov w - w'} + {\bar w\ov \bar w - \bar w'} )   + 
       \MM^2  ( {ww'\ov (w - w')^2} + {\bar w\bar w'\ov (\bar w - \bar w')^2 } ) \Big]   + O(\MM^3), \ \ \ \ 
       \ee
       where $w$ is to be replaced by \rf{50}. As a result, 
       \be \la{54} 
       \cG=   -  {1\ov   \pi}\log | z-z' |  +  {1\ov  2 \pi}  \MM  - {1\ov  4 \pi} \MM^2   { (1+ z^2) ( 1 + z'^2) \ov ( z-z')^2 }  +  O(\MM^3)\ , 
       \ee 
       where  $z,z'$   are real. 
        The   Green's function   \rf{41}   will depend on $e=L$  
       not only at the coinciding points (as is the case also  for $\MM=0$      if one uses a covariant cutoff) but also explicitly via  $\M$ or $\MM$ 
     and this makes  the 
    integral over $e$ non-trivial. This is   a reflection of  the  explicit breaking of  the conformal invariance  by the mass term at the boundary. 

       The constant $\MM$ term in \rf{54} will  not contribute to on-shell  amplitudes  so the first non-trivial correction will be at order $\MM^2$. 
       Note, however, that  significance of this expansion is unclear as we are effectively to integrate over $\MM$
       which depends on $L$ according to  \rf{6}.\foot{Formally,  the expansion in small $\MM$ can be done directly in the  path integral and will 
        correspond to  the insertion of $\int dt\ \dot x^2$  operators in addition to the  tachyon  or vector vertex operators:
        this will be  an amplitude with  an extra ``off-shell"   spin 2  open string mode vertex operators at zero momentum.}

 
\section{Remarks}

In the standard Polyakov path integral approach 
    the  integral   over  the  conformal factor $\rho$  decouples for  $D=26$ only in the vacuum partition function:
     if one considers correlators of vertex operators 
    then the     condition  of decoupling of $\rho$ leads to the mass shell  restriction on external   momenta  \ci{nepom}.
In the  ``massive"  case  discussed above  the path integral  over  the boundary value of the 
conformal  factor or  the 1d metric will  no longer be  trivial  (i.e.  will not just give,  e.g.,   the 
 delta-function  of   the on-shell tachyon scattering   condition $ \alpha' p^2_i =1$).

One   may study the  off-shell amplitudes  and hope  that looking at  their 
consistency conditions  may help determine the  modified mass shell restrictions  on the external momenta. 
This requires  extracting  
the dependence on the 1d metric   from  the  explicit factors  in the 
mass term as   well as the anomalous
contributions coming  from the use of a covariant cutoff in Green's functions at
coinciding points (giving  terms proportional to $p^2_i$).
An open  problem  is to find an approximation (e.g., small or large  mass expansion) 
that may make the study of scattering amplitudes tractable.

For example, ref.    \cite{Sever:2017ylk}   considered a  modification of   the Veneziano amplitude due to massive ($\sim m$) string   ends assuming 
that 
$ {1\ov \sqrt \a'} \ll m \ll s,t$  where $s,t$ are  kinematic variables. 
There  just the   leading semiclassical approximation  was used  assuming  that  the 
 amplitude is  still given by  the expectation value of product of $e^{i \vec p_i \vec x(t_i)}$  insertions
 (i.e. ignoring the issue  that  the structure of vertex operators  can no longer be  fixed using conformal invariance condition). 
It would be interesting to    perform a similar computation   in the setting described above where one  is supposed to integrate 
 over the 1d metric. 

\section*{Acknowledgments}

We would like to thank   J. Sonnenschein  for    discussions
  and also   M. Green  and   A. Polyakov     for  related comments. 
This work  was supported by the STFC grant ST/P000762/1.

\iffa 
 \appendix 
 \section{Comments}
   Let us consider map of a circle  to a line  e.g.  by (other options can be recalculated easily)
 \be \la{41}
 t= \tan { \vp\ov 2} \  . \ee
 Then  one finds that 
 \be \la{42} 
 G(t) = - {1\ov \pi}\log  { 2 |t| \ov \sqrt{1 + t^2} } \  , \ee 
  \be \la{43} 
 \cG(t) = G(t)  + {1 \ov 2 \pi (1 + t^2) } \Big[ (i+t)^2  \Phi \Big( { i + t \ov i-t}, 1 , 1 + \mu^{-1} \Big)
 + (i-t)^2  \Phi \Big( { i - t \ov i+t}, 1 , 1 + \mu^{-1} \Big) \Big] \ . 
 \ee 
 Expanded   in large   mass  parameter or powers of $\MM^{-1}$      this  Greens function 
  can be written as 
 \be 
  \la{44} 
 \cG(t) = {1 \ov 2 \pi} \sum_{s=1}^\infty   { 1\ov \mu^s}  \Big[ \Li_{s+1} \Big( { i + t \ov i-t}\Big) 
 +  \Li_{s+1} \Big( { i - t \ov i+t}\Big) \Big] \ .  \ee
 
 Naive expansion in small $\mu$   under the sum and then doing the sum gives 
 \be 
 \cG(t)  =  G(t)  + { 1 \ov 2 \pi} \Big[
   \mu    - {1 \ov 4}  \mu^2  (1  + t^{-2} ) + {1 \ov 8}  \mu^3  (1  + 4 t^{-2}  + 3 t^{-4} )   +  O(\mu^4) \Big] 
\la{45}
 \ee 
 
 Other   options of the map \rf{41}   are 
 \be \la{46}
  s= - \cot {\vp \ov 2} = - t^{-1}  \ , \ \ \ \ \ \ \ 
 y= 2  \sin{ \vp\ov 2 } = {  2  t  \ov \sqrt{1 + t^2} } 
 \ee
\fi 

 \iffa 
 2d conformal invariance plays a  central role because there exists a gauge where Nambu action reduces 
 to free 2d  scalar fields. But once there are boundary terms   this will  not be so as boundary terms will be breaking 
 conformal invariance. This then leads  to tension  with idea of mapping domains,  finding vertex operators etc. 
 But may be this just means that  one should not rely on  orthogonal  gauge and conformal invariance too much
 and then there is no formal issue with starting with \rf{1},\rf{2}. 
 \fi 



\begin{thebibliography}{30}
\parskip=0.2 pt



\bibitem{Chodos:1973gt} 
  A.~Chodos and C.~B.~Thorn,
  ``Making the Massless String Massive,''
  Nucl.\ Phys.\ B {\bf 72}, 509 (1974).
\bibitem{Bars:1975dd} 
  I.~Bars and A.~J.~Hanson,
  ``Quarks at the Ends of the String,''
  Phys.\ Rev.\ D {\bf 13}, 1744 (1976).
  
  
  
\bibitem{Barbashov:1977nq} 
  B.~M.~Barbashov and V.~V.~Nesterenko,
  ``Relativistic String with Massive Ends,''
  Theor.\ Math.\ Phys.\  {\bf 31}, 465 (1977).
  ``Introduction to the relativistic string theory,''
  Singapore, Singapore: World Scientific (1990),  249 p.
 V.~V.~Nesterenko,
  ``Is there tachyon in the Nambu-Goto string with massive ends?,''
  Z.\ Phys.\ C {\bf 51}, 643 (1991).

\bibitem{Sever:2017ylk} 
  A.~Sever and A.~Zhiboedov,
  ``On Fine Structure of Strings: The Universal Correction to the Veneziano Amplitude,''
  JHEP {\bf 1806}, 054 (2018)
  [arXiv:1707.05270].
  
  
  

\bibitem{Sonnenschein:2018aqf} 
  J.~Sonnenschein and D.~Weissman,
  ``Quantizing the rotating string with massive endpoints,''
  JHEP {\bf 1806}, 148 (2018)
  [arXiv:1801.00798].
  J.~Sonnenschein, D.~Weissman and S.~Yankielowicz,
  ``The scattering amplitude of stringy hadrons I: Strings with opposite charges on their endpoints,''
  arXiv:1906.00976.



\bibitem{Polyakov:1981rd} 
  A.~M.~Polyakov,
  ``Quantum Geometry of Bosonic Strings,''
  Phys.\ Lett.\  {\bf 103B}, 207 (1981).
  
\bibitem{Polyakov:1987ez} 
  A.~M.~Polyakov,
  ``Gauge Fields and Strings,''  
  Contemp.\ Concepts Phys.\  {\bf 3}, 1 (1987). Harwood Acad.\ Publ. (1987).
  
  
  
\bibitem{Fradkin:1985qd} 
  E.~S.~Fradkin and A.~A.~Tseytlin,
  ``Nonlinear Electrodynamics from Quantized Strings,''
  Phys.\ Lett.\  {\bf 163B}, 123 (1985).


 \bi{beta}
  J.~M.~F.~Labastida and M.~A.~H.~Vozmediano,
  ``Bosonic Strings in Background Massive Fields,''
  Nucl.\ Phys.\ B {\bf 312}, 308 (1989).
  I.~L.~Buchbinder, V.~A.~Krykhtin and V.~D.~Pershin,
  ``Massive field dynamics in open bosonic string theory,''
  Phys.\ Lett.\ B {\bf 348}, 63 (1995)
  [hep-th/9412132].
  ``On consistent equations for massive spin two field coupled to gravity in string theory,''
  Phys.\ Lett.\ B {\bf 466}, 216 (1999)
  [hep-th/9908028].



\bi{gr}
  D.~J.~Gross and V.~Rosenhaus,
  ``A line of CFTs: from generalized free fields to SYK,''
  JHEP {\bf 1707}, 086 (2017)
  [arXiv:1706.07015].



 \bi{alv}
  M.~Luscher, K.~Symanzik and P.~Weisz,
  ``Anomalies of the Free Loop Wave Equation in the WKB Approximation,''
  Nucl.\ Phys.\ B {\bf 173}, 365 (1980).
  E.~S.~Fradkin and A.~A.~Tseytlin,
  ``On Quantized String Models,''
  Annals Phys.\  {\bf 143}, 413 (1982).
O.~Alvarez, { ``Theory of strings with boundaries: fluctuations, topology 
and quantum geometry"}, Nucl.\ Phys.\ B {\bf 216} (1983) 125.
B. Durhuus, P. Olesen, and J.L. Petersen, 
{``On the static potential in Polyakov's theory of the quantized string"},
Nucl.\ Phys.\ B {\bf 232} (1984) 291.
  A.~G.~Cohen, G.~W.~Moore, P.~C.~Nelson, and J.~Polchinski,
  {``An off-shell propagator for string theory",}
  Nucl.\ Phys.\  B {\bf 267} (1986) 143.
  H.~Luckock,
  ``Quantum Geometry of Strings With Boundaries,''
  Annals Phys.\  {\bf 194}, 113 (1989).






\bibitem{Tseytlin:1999dj} 
  A.~A.~Tseytlin,
  ``Born-Infeld action, supersymmetry and string theory,''
  in ``Shifman, M.A. (ed.): The many faces of the superworld",  pp. 417-452
  [hep-th/9908105].


     
     
\bibitem{Schwinger:1951nm} 
  J.~S.~Schwinger,
  ``On gauge invariance and vacuum polarization,''
  Phys.\ Rev.\  {\bf 82}, 664 (1951).
  

 
  \bibitem{mob} 
  A.~A.~Tseytlin,
  ``Renormalization of Mobius Infinities and Partition Function Representation for String Theory Effective Action,''
  Phys.\ Lett.\ B {\bf 202}, 81 (1988).
  J.~Liu and J.~Polchinski,
  ``Renormalization of the Mobius Volume,''
  Phys.\ Lett.\ B {\bf 203}, 39 (1988).
  
  


\bibitem{Abouelsaood:1986gd} 
  A.~Abouelsaood, C.~G.~Callan, Jr., C.~R.~Nappi and S.~A.~Yost,
  ``Open Strings in Background Gauge Fields,''
  Nucl.\ Phys.\ B {\bf 280}, 599 (1987).
  
     \bi{pol}
      A.M.~Polyakov, unpublished (1997). 
  V.~S.~Rychkov,
  ``Wilson loops, D-branes, and reparametrization path integrals,''
  JHEP {\bf 0212}, 068 (2002)
  [hep-th/0204250].

  \bibitem{Orland:2001rq} 
  P.~Orland,
  ``Evolution of fixed end strings and the off-shell disk amplitude,''
  Nucl.\ Phys.\ B {\bf 605}, 64 (2001)
  [hep-th/0101173].
 
   
\bibitem{Makeenko:2009rf} 
  Y.~Makeenko and P.~Olesen,
  ``Wilson Loops and QCD/String Scattering Amplitudes,''
  Phys.\ Rev.\ D {\bf 80}, 026002 (2009)
  [arXiv:0903.4114].
  ``Quantum corrections from a path integral over reparametrizations,''
  Phys.\ Rev.\ D {\bf 82}, 045025 (2010)
  [arXiv:1002.0055].
  ``Scattering amplitudes of QCD string in the worldline formalism,''
  Phys.\ Part.\ Nucl.\  {\bf 45}, 771 (2014)
  [arXiv:1208.1209].
 J.~Ambjørn, Y.~Makeenko and A.~Sedrakyan,
  ``Effective QCD string beyond the Nambu-Goto action,''
  Phys.\ Rev.\ D {\bf 89}, no. 10, 106010 (2014)
  [arXiv:1403.0893].
   
  
  
  \bi{nepom}
  R.~I.~Nepomechie,
  ``Duality of the Polyakov $N$ Point Amplitude,''
  Phys.\ Rev.\ D {\bf 25}, 2706 (1982).
  H.~Aoyama, A.~Dhar and M.~A.~Namazie,
  ``Covariant Amplitudes in Polyakov's String Theory,''
  Nucl.\ Phys.\ B {\bf 267}, 605 (1986).
  
  
   

  
\end{thebibliography}
\end{document}

\bibitem{Polyakov:2000fk} 
  A.~M.~Polyakov,
  ``String theory as a universal language,''
  Phys.\ Atom.\ Nucl.\  {\bf 64}, 540 (2001)
  [Int.\ J.\ Mod.\ Phys.\ A {\bf 16}, 4511 (2001)]
  [hep-th/0006132].

\ed

\